\documentclass{elsart}

\usepackage{natbib}

 \usepackage{graphics}
 \usepackage{graphicx}
 \usepackage{epsfig}

\usepackage{amssymb}

\begin{document}

\newcommand{\pam}{\textsf{PAMELA}}

\begin{frontmatter}



\title{Detection of the high energy   component of Jovian electrons in Low Earth Orbit with the PAMELA experiment.}


\author[a]{M.~Casolino\corauthref{cor}},
\corauth[cor]{Corresponding author}
\ead{Marco.Casolino@roma2.infn.it}
\author[a]{N.~De Simone},
\author[a]{V.~Di Felice},
\author[a]{P.~Picozza}
\author{On Behalf of the PAMELA collaboration}

 \address[a]{INFN, Structure of Rome II, and Physics Department,
  University of Rome II ``Tor Vergata'', I-00133 Rome, Italy}

\begin{abstract}
% Text of abstract
The   \pam\ experiment is devoted to the study of cosmic rays in
Low Earth Orbit with an apparatus optimized to perform a precise
determination of the galactic antimatter component of c.r. It is
constituted by a number of detectors built around a permanent
magnet spectrometer. \pam\ was launched in space on June $15^{th}$
2006 on board the Russian Resurs-DK1 satellite for a mission
duration of three years. The characteristics of the detectors, the
long lifetime and  the orbit  of the satellite, will allow to
address several aspects  of cosmic-ray physics. In this work we
discuss the observational capabilities of \pam\ to detect the
electron component above 50 MeV. The magnetic spectrometer allows
a detailed measurement of the energy spectrum of electrons of
galactic and Jovian origin. Long term measurements and
correlations with Earth-Jupiter 13 months synodic period will
allow to  separate these two contributions and to measure the
primary electron  Jovian component, dominant  in the 50-70 MeV
energy range. With this technique it will also be possible to
study the contribution  to the electron spectrum  of Jovian $e^-$
reaccelerated up to 2 GeV at the Solar Wind Termination Shock.
\end{abstract}

\begin{keyword}
Cosmic rays \sep Satellite-borne experiment \sep Solar wind  \sep Jovian electrons
\PACS 96.40.-z \sep 07.87.+v \sep 96.60.Vg \sep 96.30.Kf

\end{keyword}
{\small Accepted for publication on Advances in Space Research,
http://dx.doi.org/10.1016/j.asr.2007.07.024}

\end{frontmatter}

% main text
\section{Introduction}
\label{sec:overview}   \pam\   is a satellite-borne apparatus
devoted to the study of cosmic rays, with an emphasis on its
antiparticle component. The satellite  was launched on June the
$15^{th}$ 2006 from the cosmodrome of Baikonur with a Soyuz
rocket. The Resurs-DK1 satellite housing \pam\ on one side flies
on a quasi-polar (inclination 70$^{\circ}$), elliptical (altitude
350--600~km) orbit  with an expected mission duration of 3
years\cite{pamorbit}. Taking advantage of the orbital
characteristics of the satellite, its long observational lifetime,
and   the structure of the detector, the mission will be able to
address several aspects of cosmic-ray physics  over an energy
range not previously reached by balloon-borne experiments. The
results will increase our knowledge of cosmic ray origin and
propagation, as well as shed some light on some cosmological
questions. In the field of heliospheric physics it will be
possible to study solar modulation and solar particle events over
an energy range  and with a precision insofar not measured in
orbit\cite{pamhelio}. In this work we evaluate the observational
capabilities of \pam\ in respect to electrons of Jovian origin.
\pam\ is sensitive to jovian electrons in the energy range 50-70
MeV, where they are dominant over the galactic component. At
higher energies it should be possible to identify the component
reaccelerated at the Solar Wind Termination Shock, up to $\simeq 2
$ GeV by extracting the component  modulated over the
Earth-Jupiter synodic year of 13 months.

\subsection{The \pam\ instrument}

\begin{figure}

\begin{center}
\includegraphics[scale=0.60]{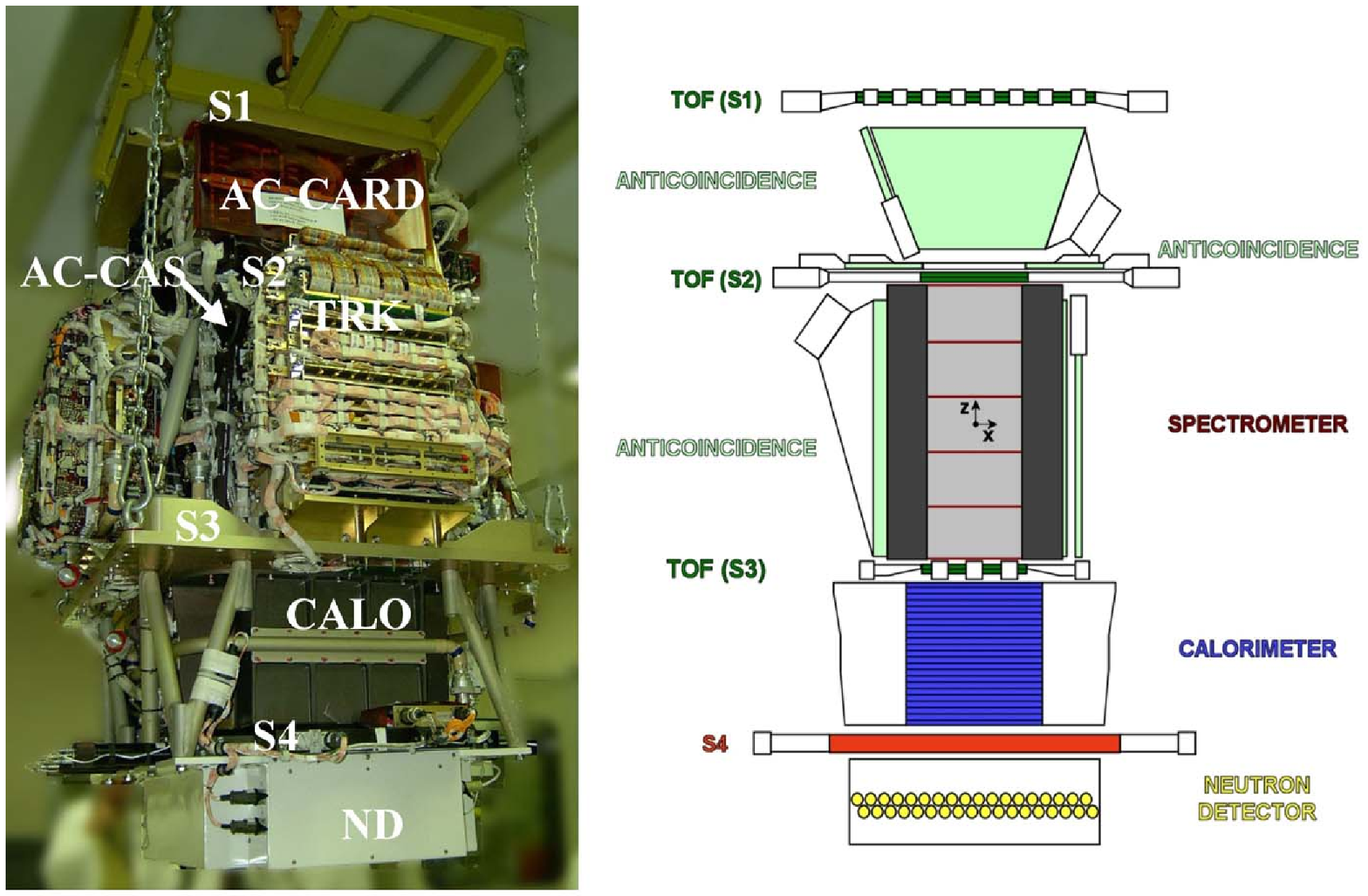}
 \caption{Left: Photo of
the  \pam\ detector during the final integration phase in Tor
Vergata clean room facilities, Rome. It is possible to discern,
from top to bottom,  the topmost scintillator system, S1,  the
electronic crates around the magnet spectrometer, the baseplate
(to which \pam\ is suspended by chains), the black structure
housing the Si-W calorimeter, S4 tail scintillator and the neutron
detector.  Right: scheme - approximately to scale with the picture
-  of  the detectors composing \pam. } \label{scheme}
\end{center}
\end{figure}

\pam\ is constituted by a number of highly redundant detectors
capable of identifying  particles providing   charge, mass,
rigidity and velocity information  over a very wide energy range.
The instrument (see Figure \ref{scheme}) is built around a
permanent magnet with a silicon microstrip   tracker. A
scintillator system provides trigger, charge and time of flight
information; a silicon-tungsten calorimeter is used to perform
hadron/lepton separation. A shower tail catcher and a neutron
detector at the bottom of the apparatus increase the particle
identification capability. An anticounter system is used to reject
spurious events in the off-line phase. Around the detectors are
housed the readout electronics, the interfaces with the CPU and
all primary and secondary power supplies. All systems (power
supply, readout boards etc.)  are redundant with the   exception
of the CPU \cite{cpu} which is more tolerant to failures. The
system is enclosed in a pressurized container located on one side
of the Resurs-DK1 satellite. In a twin  pressurized  container is
housed the   Arina experiment, devoted to the study of the low
energy trapped electron and proton component. Total weight of
\pam\ is 470 kg;  power consumption is 355 W, geometrical factor
is 21.5 $cm^2 sr$. The satellite flies on a quasi-polar
(inclination 70$^{\circ}$), elliptical (altitude 350--600~km)
orbit  with an expected mission length of 3 years. Here we briefly
describe   the main characteristics of \pam\ subdetectors; a more
detailed description of the device can be found in
\cite{pamela,pamgene}.
\begin{itemize}
\item{\bf Scintillator / Time of Flight}
The scintillator system~\cite{tof} provides trigger for the
particles  and time of flight information for incoming particles.
There are three scintillators layers, each composed by two
orthogonal   planes  divided in various bars (8 for S11, 6 for
S12, 2 for S21 and S22 and 3 for S31 and S32) for a total of 6
planes  and 48 phototubes (each bar is read by two phototubes). S1
and S3 bars  are 7 mm thick and  S2 bars are 5 mm. Interplanar
distance between  S1-S3 of  77.3 cm allows for a TOF determination
of
 250
 ps precision for protons and 70 ps for C nuclei (determined with  beam tests in GSI), allowing separation of electrons from antiprotons up to $\simeq 1$ GeV and albedo rejection. The scintillator system is also capable of providing charge information up to $Z=8$.
\item{\bf Magnetic Spectrometer}
The permanent magnet is composed of 5 blocks, each divided in 12
segments of  Nd-Fe-B alloy with a residual magnetization of 1.3 T
arranged to provide an almost uniform magnetic field along the $y$
direction. The size of the cavity is $13.1\times 16.1\times 44.5\:
cm^3$, with a mean magnetic field of 0.43 T. Six layers of $300\:
\mu m$ thick double-sided microstrip silicon detectors~\cite{spec}
are used to measure particle deflection with $3.0\pm 0.1\: \mu m$
and $11.5\pm 0.6 \: \mu m$ precision in the bending and
non-bending views. Each layer is made  by three ladders, each
composed by two  $5.33\times 7.00\: cm^2$ sensors coupled to a VA1
front-end hybrid circuit. Maximum Detectable Rigidity (MDR) has
been measured on CERN proton beam and found $\simeq 1$ TV.

\item{\bf Silicon Tungsten Calorimeter}
Lepton/Hadron discrimination is performed by the Silicon Tungsten
sampling calorimeter~\cite{calo} located on the bottom of \pam\ .
It is composed of 44 silicon layers  interleaved by 22  0.26 cm
thick tungsten plates. Each silicon layer is composed arranging
$3\times 3$ wafers, each of $80\times 80\times 0.380\: mm^3$ and
segmented in 32 strips, for a total of 96 strips / plane. 22
planes are used for  the X view and 22  for  the Y view in order
to provide topological and energetic information of the shower
development in the calorimeter. Tungsten was chosen in order to
maximize electromagnetic radiation  lenghts (16.3 $X_o$)
minimizing hadronic interaction lenght (0.6 $\lambda $). The
CR1.4P ASIC chip is used for front end electronics, providing a
dynamic range of 1400 mips (minimum ionizing particles) and
allowing nuclear identification up to Iron.
\item{\bf Shower tail scintillator}
This 1 cm thick scintillator ($48\times 48\: cm^2$  wide) is
located below the calorimeter and is used to improve hadron/lepton
discrimination by measuring the energy not contained in the shower
of the calorimeter. It can also function as a standalone trigger
for the neutron  detector.
\item{\bf Neutron Detector}
The $60 \times  55 \times 15\:\: cm^3$ neutron  detector~\cite{nd}
is composed by 36 $^3He$ tubes arranged in two layers and
surrounded by polyethylene shielding and a 'U' shaped cadmium
layer to remove thermal neutrons not coming from the calorimeter.
It is used to improve hadron/lepton identification by detecting
the number of neutrons produced in the hadronic and
electromagnetic cascades. Since the former have a much higher
neutron cross section than the latter, where neutron production
comes essentially through a nuclear  giant resonance, it is
estimated that \pam\ overall identification capability is improved
by a factor 10. As already mentioned, the neutron detector is used
to measure neutron field in Low Earth Orbit (LEO) and in case of
solar particle events, as well as in  the high energy lepton
measurement.
\item{\bf Anticoincidence System}
To reject  spurious triggers due to interaction with the main body
of the satellite, \pam\ is shielded by a number of scintillators
used with anticoincidence functions~\cite{anti}. CARD
anticoincidence system is composed of four 8 mm thick
scintillators  located in the area between S1 and S2. The Top
Anticounter (CAT) is a  scintillator placed on top of the magnet:
it is composed by a single piece with a central hole where the
magnet cavity is located and  read out by 8 phototubes. Four
scintillators, arranged on the sides of the magnet, make the side
(CAS) lateral anticoincidence system.

\end{itemize}

\section{Jovian electrons}

\label{sec:Jovian} Since the discovery
  of Jovian electrons  of $1.75 MeV \leq E \leq 25 MeV$  at about 1~AU from
Jupiter by   Pioneer 10~\cite{simpson, eraker},  several
interplanetary missions have measured this component of cosmic
rays in different points of the heliosphere. At 1~AU from the Sun
the IMP-8 satellite could detect Jovian electrons in the range
between 0.8 and 16~MeV and measure their 27 days modulation by the
passage of Coronal Interaction Regions
(CIR)~\cite{eraker,chenette}. Measurements performed with the
electron spectrometer on board ISEE spacecraft  in the energy
range 5-30 MeV over 6 synodic periods during solar
maximum~\cite{moses} and obtaining a power law index $\simeq 1.5$
up to 10 MeV and steepening to $ \leq 6$ above. Most prominent
feature of the Jovian flux is the long term modulation of 13
months related to the Earth-Jupiter synodic year. Since Jovian
electrons propagate along  the interplanetary magnetic field
lines, when the two planets are on the same solar wind spiral
line, the electron transit from Jupiter to the Earth is eased and
flux increases. When the two planets lie on different spiral lines
the electron flux decreases. Currently we know that Jupiter is the
strongest electron source at low energies (below 25~MeV) in the
heliosphere within a radius of 11~AUs from the Sun. Measurement of
the power law spectrum up to $\approx 10 ~MeV$ with EPHIN
instrument on board SOHO spacecraft confirmed a  power law
spectrum with  spectral index $\gamma$ = 1.65 in maximum jovian
flux days and $\gamma$ = 1.51 in minimum jovian flux
\cite{ephinsoho1,ephinsoho,gomez}. Ulysses has performed detailed
measurements in a wide range of heliographic latitudes and at
various distances from Jupiter studying the geographical and
temporal variability of the diffusion coefficients
\cite{heber,heberrev} and the    high energy (up to 9 GeV)
galactic electron component \cite{ferrando}. Assuming those
spectral indexes  \cite{potgieter} estimated  the galactic
component to become  dominant above the primary Jovian component
above $\simeq 70$ MeV.

Many models attempt to describe the complex processes of Jovian
and galactic electron diffusion in the solar wind. For instance,
in \cite{chenette} a convection diffusion model is used applying
the results to IMP8 and Pioneer 10 and 11 data. More recently,
\cite{Ferreira2001a} and \cite{Ferr2005} developed a three
dimensional  model with convection, gradient and curvature drift
and current sheet effects. Measurements in the energy range 50-130
MeV at Earth will contribute  to test the validity of these models
at high energies and in the recovery period from solar minimum. It
is known that  cosmic rays originating outside the heliosphere can
be accelerated at the solar wind termination shock
\cite{jk91,mor1999,potgieter}. This applies also to Jovian
electrons, which are transported outward by the solar wind, reach
the TS and   undergo shock acceleration thus increasing their
energy. Some of these electrons are scattered back in the
heliosphere and reach  Earth. Reacceleration intensity and
spectral feature depends upon shock position\cite{potgieter,pot2}:
a precise measure of the high energy component and its modulation
with synodic year could also contribute to determine the overall
characteristics of the shock.

\section{\pam\ observational capabilities}

Electrons can be detected by \pam\ in the energy range between 50
MeV and 400 GeV measuring their energy with the magnetic
spectrometer. The spectrometer allows also separation and
identification of the positron component in the 50 MeV - 290 GeV
range.  The electron spectrum  has  been measured at Earth  by
several space-borne (e.g.
%ISEE-3 (7-30MeV) \cite{moses}
OGO-V \cite{meyer} $\sim 10-200\: MeV$) and  balloon-borne
experiments (e.g. \cite{evenson,boe00},   $\gtrsim 30 $GeV). A
precise determination of the electron and positron spectra  and
the  temporal variation during recovery from the solar minimum
will allow to   gather information on solar modulation, also in
respect to  charge dependent \cite{jt,bess}   effects, reducing
the systematics constraining propagation  models in the galaxy and
in the solar system. To determine the observational capabilities
of \pam\ in the range 50 MeV to 2 GeV we have divided the particle
spectra in different ranges according to the spectral shape of
galactic and Jovian electrons:
\begin{itemize}
\item 50--70~MeV: {\it non-reaccelerated component of Galactic and Jovian $e^-$}.
The electrons in this range, at the lower limit of
  \pam\ detection capabilities, represent the primary
  non-reaccelerated component. These electrons are mostly of Jovian origin and do not undergo acceleration
  at the TS. Their long and short
   term   modulation   would give information on high energy acceleration phenomena in
   the Jovian magnetosphere and on propagation effects in the inner    heliosphere.
\item 70--130~MeV: {\it accelerated component of Galactic $e^-$, primary  Jovian $e^-$. }
This is the highest energy   electrons are believed to be accelerated by Jupiter. In this
range Jovian flux is less abundant than Galactic one, but it will be possible to extract this signal
 thanks to its  synodic modulation.
\item 130--600~MeV: {\it accelerated component of Galactic and  Jovian $e^-$ toward the maximum.} In this energy range Jovian electrons are  reaccelerated at the TS.
 \item 600~MeV--2 GeV:{\it accelerated component of Galactic and  Jovian $e^-$ from  the maximum.}
 These    allows to gather a large number of events of the high energy component of electrons of Jovian origin. 2 GeV has been taken as the maximum detectable energy for
accelerated electrons according to \cite{potgieter}.  In this and
the former case it will be possible to extract the signal assuming
the presence of  synodic year  modulation  at these energies.
\end{itemize}

Given a differential   electron flux $\phi(E)$  in an energy range
$[E_1, E_2]$  with an instrument with geometrical factor $G(E)$
and live time T(E) due to energy dependent   cutoff, the expected
number of electrons N is given by:
\begin{equation}
\label{eq:Nintegrale} N =
\int^{E_2}_{E_1}\phi\left(E\right)G\left(E\right)T\left(E\right)\;dE
\end{equation}

assuming an efficiency 1 of the detector. \pam\ minimum threshold
energy for electron detection is 50~MeV. Below this energy
particles are completely  deflected by the magnetic field of the
permanent magnet and do not reach the bottom scintillator (S3). At
this threshold energy the geometrical factor of \pam\ is   equal
to $G_0= 1.4$ $cm^2 sr$ due to the fact that only a small
percentage of particles can reach S3 to give a valid trigger for
the apparatus. The energy dependent geometrical factor has been
approximated with the following formula \cite{ricciarini}:
\begin{equation}
\label{eq:rette1}
G\left(E\right)=\left\{\begin{array}{ll}  \alpha E + G_0 & E<130~ MeV \\
                                                                                        aE^b & 130<E<600~ MeV \\
                                                                                        21.5\:  cm^2 sr &  E>600~ MeV\\
                                                                                        \end{array} \right.
\end{equation}
Also geomagnetic shielding reduces total counts, allowing low
energy particles to be acquired only in high latitude regions
where cutoff is lower. The amount of time T(E) that \pam\ spends
in locations accessible to electrons  of energy E or higher has
been evaluated with the Stormer cutoff approximation and linearly
approximated in the energy range of interest:
\begin{equation}
T\left(E\right)=\beta E + T_0
\end{equation}

The galactic and Jovian fluxes can be approximated by power law spectra of the form:
\begin{equation}
\label{eq:Ndifferenziale}
\phi\left(E\right)=N_1\left(\frac{E}{E_1}\right)^\gamma
\end{equation}
in  Table \ref{tab:flusso} are shown the flux and spectral indexes
evaluated  from \cite{potgieter} assuming a TS at 90~AU, and
Heliospheric boundary at 120~AU. The integral \ref{eq:Nintegrale}
can be calculated analitically, resulting in:
\begin{eqnarray}
\label{eq:integralesolved1}
\small N=-N_1 E_1\left\{\frac{\alpha\beta}{\gamma+3}\left[1-\left(\frac{E_2}{E_1}\right)^{\gamma + 3}\right]E_1^2+ \right.\\ +\left. \frac{\left(\alpha T_0+\beta G_0\right)}{\gamma + 2}\left[1-\left(\frac{E_2}{E_1}\right)^{\gamma + 2}\right]E_1+\frac{G_0 T_0}{\gamma + 1}\left[1-\left(\frac{E_2}{E_1}\right)^{\gamma + 1}\right]\right\}
\end{eqnarray}
 for $E<130$ MeV and $E>600$ MeV  and

 \begin{equation}
        N=\frac{N_1 a \beta}{E_1^\gamma \left(\gamma + b + 2 \right)}\left[E_2^{\gamma+b+2}-E_1^{\gamma+b+2}\right]
        +\frac{N_1 a T_0}{E_1^\gamma \left(\gamma + b + 1 \right)}\left[E_2^{\gamma+b+1}-E_1^{\gamma+b+1}\right]
        \end{equation}
        for $130<E<600$ MeV.

In Table \ref{tab:ResultsFluxJovian2} are shown the expected  counts for the all particle flux and the Jovian component.
Aside from the 50-70 MeV range, where Jovian component is dominant,
 to separate the
two components at higher energies it will be necessary to gather
statistics over a time of the order of some  months assuming that
this signal will be modulated by Earth-Jupiter relative
position\footnote{The two signals can also be separated using the
maximum likelihood method  assuming a sinusoidal signal with 13
months periodicity.}. It is   possible to see how, with \pam , it
will be possible to study for the first time the high energy
Jovian electron component and measure the intensity of
reacceleration at the solar wind termination shock  and how this
component is affected by the Earth-Jupiter synodic year.

\section{Conclusions}

In this work we have   described  the possibilities of \pam\ to
observe electrons of Jovian origin.  This will be the first time a
magnetic spectrometer telescope in low Earth orbit will be
operational for long duration observation. Thus, it will   be
possible to perform direct measurements of both the primary and
reaccelerated electron component  extracting  the reaccelerated
Jovian  component from the galactic flux and studying  the effect
of modulation due to synodic year.
   In addition to these
phenomena, charge dependent modulation effects will be studied by
comparing the temporal dependence of electron and positron
spectra.

% \appendix

% \section{}
% \label{}

% Bibliographic references with the natbib package:
% Parenthetical: \citep{Bai92} produces (Bailyn 1992).
% Textual: \citet{Bai95} produces Bailyn et al. (1995).
% An affix and part of a reference:
%   \citep[e.g.][Ch. 2]{Bar76}
%   produces (e.g. Barnes et al. 1976, Ch. 2).

\begin{table}
    \begin{center}
        \begin{tabular}{|c|c|c|c|c|c|c|}
            \hline

        \footnotesize   $\left[E_1,E_2\right]$ & \footnotesize  $\alpha  $ &\footnotesize $G_0 $ & \footnotesize$\beta\ \ $ & \footnotesize $T_0 $ & \footnotesize $a\ $ & \footnotesize $b$\\
          $\left(GeV\right)$  & $ \left(m^2\:sr\:GeV^{-1}\right)$ & $ \left(m^2\:sr\right)$ &  $\left(s/GeV\right)$ & $\ \ \left(s\right)$ & $\left(cm^2\:sr\right)$ &  $\left(\ MeV\right)$ \\
            \hline \hline
            \footnotesize 0.05 - 0.07 &  \footnotesize $3.4\cdot 10^{-2}$ &\footnotesize  $1.4\cdot10^{-4}$ & \footnotesize $2.5\cdot10^{4}$ & \footnotesize $3.4\cdot10^{3}$ & \footnotesize   &\footnotesize \\
        \footnotesize 0.07 - 0.13 & \footnotesize   $1.5\cdot10^{-2}$ & \footnotesize $-3.5\cdot10^{-4}$ & \footnotesize $2.5\cdot10^{4}$ & \footnotesize $3.4\cdot10^{3}$ & \footnotesize & \footnotesize\\
    \footnotesize   0.13 - 0.60 & \footnotesize & \footnotesize & \footnotesize $2.1\cdot10^{4}$ & \footnotesize $4.1\cdot10^{3}$ & \footnotesize 23 & \footnotesize 76\\
        \footnotesize 0.60 - 2.0 &  \footnotesize   & \footnotesize $2.1\cdot10^{-3}$ & \footnotesize $1.1\cdot10^{4}$ & \footnotesize $1.2\cdot10^{4}$ &\footnotesize & \footnotesize\\
        \hline
        \end{tabular}
        \normalsize
    \end{center}
    \vspace{.5cm}
    \caption{Fitting parameters for geometrical factor G(E)  ($\alpha$, $G_0$, a, b), and live time T(E) ($\beta$, $T_0$) for the different energy ranges  considered (see text).}
    \label{tab:totali}
\end{table}

\begin{table}
    \begin{center}
        \begin{tabular}{|c|c|c|c|c|}
            \hline
            \footnotesize $\left[E_1,E_2\right]\ $ & \footnotesize $N_1(gal)\ \ $ &   $\gamma_{gal} $ & \footnotesize $N_1(jov)  $ & $\gamma_{jov}$   \\
              $\left(GeV\right)$ & $\left(m^2 s\:sr\:GeV\right)^{-1}$& & $\left(m^2 s\:sr\:GeV\right)^{-1}$ &  \\
            \hline \hline
    \footnotesize   0.05 - 0.07 & \footnotesize 10 & \footnotesize -2.5& \footnotesize10 & \footnotesize -3.42 \\
    \footnotesize   0.07 - 0.13 & \footnotesize 4 & \footnotesize 1.38 & \footnotesize 3.3 & \footnotesize -3.42 \\
\footnotesize       0.13 - 0.60 & \footnotesize 12.5 & \footnotesize 1.38 & \footnotesize 0.143 & \footnotesize 0.98 \\
        \footnotesize 0.60 - 2.0 & \footnotesize 60.5 & \footnotesize -2.18 & \footnotesize 0.6 & \footnotesize -2.8 \\
        \hline
        \end{tabular}
        \normalsize
    \end{center}
    \vspace{.5cm}
    \caption{Particle flux and spectral indexes  assumed to estimate  galactic and Jovian fluxes in the different energy ranges. }
    \label{tab:flusso}
\end{table}

\begin{table}
    \begin{center}
        \begin{tabular}{|c|c|c|c|c|}
        \hline
    $\left[E_1,E_2\right]\ \ $ & $N_{tot}$ & $N_{J}$  & Min. Sampl. & Percent. \\
    $\left(GeV\right)$ & $(month^{-1})$  & $(month^{-1})$ &    time (months)&  signal \\

        \hline \hline
      0.05 - 0.07 & $10\pm 3$ & $9\pm 3$  & 1   & 90$\%$\\
    0.07 - 0.13 &   $86\pm 9$  & $12\pm 3$ & 5   & 14$\%$\\
        0.13 - 0.60 & $(220\pm 2)\cdot 10^2$& $(150\pm 12)$  & 5   & 0.7$\%$\\
        0.60 - 2.0 & $(339\pm 2)\cdot 10^2$ & $(250\pm 16)$  & 4   & 0.7$\%$\\
        \hline
        \end{tabular}
    \end{center}
    \vspace{.5cm}
    \caption{Expected   counts for all particle and Jovian  flux according to equation \ref{eq:integralesolved1}.
    The sampling time shows the minimum time required to get a signal at a $2\sigma$ level  above galactic background, assuming statistical errors.  }
    \label{tab:ResultsFluxJovian2}
\end{table}

\begin{thebibliography}{elsart-harv}

  \bibitem[Asaoka 2002]{bess}
Asaoka,  Y.~, Shikaze,  Y., Abe, K., et al., \emph{ Measurements
of Cosmic-Ray Low-Energy Antiproton and Proton Spectra in a
Transient Period of Solar Field Reversal}, Phys. Rev. Lett.  88,
051101-051104, 2002; astro-ph/0109007.
\bibitem[Bonechi et al. 2007]{spec}Bonechi, L., Adriani, O., Bongi, M.,  et al., \emph{Status of the PAMELA silicon tracker}, NIM A, {\bf 570},
2  281-285, 2007.


  \bibitem[Boezio et al. 2000]{boe00}
Boezio, M., Carlson, P., Francke, T.,  et al., \emph{ The
Cosmic-Ray Electron and Positron Spectra Measured at 1 AU during
Solar Minimum Activity}, Astrophys. J. 532, 653-669, 2000.

\bibitem[Boezio et al. 2002]{calo} Boezio, M., Bonvicini, V., Mocchiutti, E., et al., \emph{ A High Granularity
  Imaging Calorimeter for Cosmic-Ray Physics}, Nucl. Instr. and
  Meth. in Phys. Res. A 487, 407-422, 2002.

\bibitem[Casolino et al. 2006]{cpu}  Casolino, M., Altamura, F., Basili, A., et al., \emph{ The PAMELA Storage and Control Unit
}, Adv. Sp., 37, 10, 1857-1861, 2006.
\bibitem[Casolino et al. 2007]{pamgene} Casolino, M., Picozza, P., Altamura, F.,  et al., {\emph Launch of the Space experiment
\pam }, in press on Adv. Space Res.,
http://dx.doi.org/10.1016/j.asr.2007.07.023,
http://arxiv.org/abs/0708.1808v1
\bibitem[Casolino and Picozza 2007]{pamorbit} Casolino, M., Picozza, P.,  {\emph Launch and Commissioning of the  \pam\ experiment on board the Resurs-DK1
satellite}, in press on Adv. Space Res.,
http://dx.doi.org/10.1016/j.asr.2007.06.062
\bibitem[Casolino and Picozza  2007b]{pamhelio} Casolino, M., Picozza,  P., {\emph The PAMELA experiment: a spaceborne observatory for heliospheric
phenomena}, in press on Adv. Space Res.,
http://dx.doi.org/10.1016/j.asr.2007.06.061

\bibitem[Chenette 1980]{chenette}Chenette, D.L., \emph{ The propagation of
  Jovian electrons to earth}, Jour. Geophys. Res. 85, 2243-2256, 1980.
\bibitem[Eraker 1982]{eraker}Eraker, J.H., \emph{ Origins of the low-energy
  relativistic interplanetary electrons}, Astrophys. Jour. 257, 862-880,
  1982.
      \bibitem[del Peral et al. 2002]{ephinsoho1} del Peral, L.,  Gomez-Herrero, R.,   Rodriguez-Frias, et al., \emph{Jovian electrons during solar quiet periods}, proc. of ''SOLSA: The Second Solar Cycle
   and Space Weather Euroconference'', Vico Equense, Italy, 24-29 Sept. 2001, ESA SP-477, 335-338, 2002.

  \bibitem[del Peral et al. 2003]{ephinsoho} del Peral, L.,  Gomez-Herrero, R.,   Rodriguez-Frias,   et al.,  \emph{Jovian electrons in the heliosphere: new insights
from EPHIN on board SOHO}, Astrop. Phys. 20,  235–245, 2003.

  \bibitem[Evenson et al. 1983]{evenson} Evenson, P., Garcia-Munoz, M., Meyer, P., Pyle, K. R., Simpson, \emph{ J. A.
    A quantitative test of solar modulation theory - The proton, helium, and electron spectra from 1965 through 1979},
    Astrophys. J., 275, L15-18, 1983.
\bibitem[Ferrando 1997]{ferrando} Ferrando, P., \emph{ MeV to GeV
electron propagation and modulation: Results of the KET  telescope
on board Ulysses}, Adv. Space Res.,  {\bf 19}, 6, 905-915, 1997.
 \bibitem[Ferreira et al. 2001]{Ferreira2001a} Ferreira, S. E. S.,
 Potgieter, M. S., Burger, R. A., Heber, B., Fichtner, H.,
 \emph{Modulation of Jovian and galactic electrons in the heliosphere 1.
 Latitudinal transport of a few MeV electrons}, J. Geophys. Res, Vol. 106, A11, 24979-24987 2001.
\bibitem[Ferreira 2005]{Ferr2005} Ferreira S. E. S., \emph{ The transport of galactic and Jovian
cosmic ray electrons in the heliosphere}, Adv. Space Res., 35,
586-596 2005.
\bibitem[Fichtner et al. 2001]{pot2} Fichtner,  H., Potgieter, M. S., Ferreira,  S. E. S., Heber, B., and
Burger, R. A.,\emph{Time-dependent 3-D modelling of the
heliospheric propagation of few-MeV electrons},
 Proc. 27$^{\mathrm{th}}$ ICRC, Hamburg,  SH 3666-3670,   2001.
\bibitem[Galper et al. 2001]{nd} Galper, A.M., et al., \emph{Measurement of primary
  protons and electrons in the energy range of 10$^{11}$--10$^{13}$ eV
  in the PAMELA experiment}, Proc. 27$^{\mathrm{th}}$ ICRC,
  Hamburg,2219-2222,   2001.
  \bibitem[Gomez-Herrero et al. 2001]{gomez} Gomez-Herrero, R.,  Rodriguez-Frias, M. D., del Peral, L.,
  Sequeiros, J.,
  Muller-Mellin, R., Kunow,  H., \emph{ Heliospheric electrons from
  Jupiter}, Proc. 27$^{\mathrm{th}}$ ICRC, Hamburg,    SH 3601-3604, 2001.
\bibitem[Heber et al. 2002]{heber} Heber, B., Ferrando P.,
Raviart A.,  et al, \emph{ 3 – 20 MeV electrons in the inner
three-dimensional heliosphere: Ulysses Cospin/Ket observations.}
Astrophys. Journ., {\bf 579},  888-894, 2002.

  \bibitem[Heber et al. 2007]{heberrev} Heber, B., Potgieter M.
  S., Ferreira, S.E.S, et al, \emph{ An overview of Jovian
  electrons during the distant Ulysses Jupiter flyby}, Plan. and
  Space Science, {\bf 55}, 1-11, 2007.
  \bibitem[Jokipii and Thomas  1981]{jt}
    Jokipii, J. R.; Thomas, B.,  \emph{ Effects of Drift on the Transport of Cosmic Rays: IV. Modulation by a Wavy
Interplanetary Current Sheet}, Astrophys. Journ.,  {\bf 243},
1115-1122,  1981.
\bibitem[Jokipii and Kota 1991]{jk91} Jokipii, J.R., Kota, J. \emph{ On the interpretation of the high
cosmic ray electron fluxes observed in 1986}, Proc.  of the 22nd
ICRC, 3, 569-572, 1991.
\bibitem[L'Heureux and Meyer 1976]{meyer}
L'Heureux J. and Meyer P., \emph{ Quiet-time increase of low
energy electrons: The Jovian origin}, J. Astrophys., 209 955-960,
1976.

\bibitem[Moraal et al. 1999]{mor1999} Moraal, H., Steenberg, C.D., Zank, G.P., \emph{  Simulations of galactic and
anomalous cosmic ray transport in the heliosphere}, Adv. Space
Res.,  23 (3), 425–-428, 1999.
\bibitem[Moses 1987]{moses}Moses,  D., \emph{Jovian Electrons at 1 AU:
1978-1984}, Astrophys. Journ., {\bf 313}, 471-486, 1987.
\bibitem[Orsi et al. 2006]{anti} Orsi, S., et al., \emph{ The anticoincidence shield of the PAMELA space experiment },  Advances in Space Research, {\bf  37},  10, 1853-1856, 2006.

\bibitem[Osteria  et al. 2004]{tof}Osteria, G., Barbarino, G.C. et al., \emph{
 The ToF and Trigger electronics of the PAMELA experiment}, NIM A,
 {\bf  518}, 1-2, 161-163, 2004.
\bibitem[Picozza et al. 2007]{pamela} Picozza, P., Galper, A. M., Castellini, G., et al.,\emph{  PAMELA - A Payload for Antimatter Matter Exploration and Light-nuclei
Astrophysics},  Astrop. Phys., {\bf 27}, 4, 296-315, 2007.
astro-ph/0608697.
\bibitem[Potgieter and Ferreira 2002]{potgieter} Potgieter, M.S.  and
  Ferreira, S.E.S., \emph{ Effects of the solar wind termination shock on the modulation of Jovian and galactic electrons in the heliosphere},
  Jour. Geophys. Res. 107  A7, 1089, 2002.
\bibitem[Ricciarini 2005]{ricciarini} Ricciarini, S., \emph{ Development of tracking system acquisition electronics
 and analysis of first data for the PAMELA experiment},
             PhD thesis, Universita' degli Studi, Firenze, Italy, 2005.
\bibitem[Simpson et al. 1974]{simpson} Simpson, J.A., Hamilton, D.,  Lentz, G.,  et al., \emph{ Protons and
  Electrons in Jupiter's Magnetic Field: Results from the University
  of Chicago Experiment on Pioneer 10}, Science 4122, 306-309, 1974.
\end{thebibliography}
\end{document}